\documentclass{ledger}

\usepackage{xcolor}

\newcommand{\thefirstpagenum}[0]{1}
\newcommand{\thelastpagenum}[0]{5}
\newcommand{\theyear}[0]{201X}
\newcommand{\thevol}[0]{X}
\newcommand{\thedoi}[0]{DOI 10.5195/LEDGER.\theyear.X}
\newcommand{\ledgerpages}[0]{\thefirstpagenum-\thelastpagenum}

\usepackage{tkz-kiviat,numprint} 
\usetikzlibrary{arrows}
\usepackage{hyperref}

\usepackage{graphicx}

\usepackage{todonotes}

\hypersetup{pdfauthor={Jan Gorzny}, pdftitle={Classifying Rollups}}

\title{A Rollup Comparison Framework}
\author{Jan Gorzny\thanks{Jan Gorzny (jan@zircuit.com) is a Technical Lead at Zircuit} \and Martin Derka\thanks{Martin Derka (martin@zircuit.com) is a Technical Lead at Zircuit}}

\pretitle{
  \centering \selectfont  RESEARCH ARTICLE \par 
  \fontsize{24pt}{28pt}\selectfont} 

\begin{document}

\maketitle

\thispagestyle{pagefirst}

\begin{abstract}
Rollups are a popular blockchain paradigm where one blockchain network is anchored to a different blockchain network, typically though smart contracts and data commitments.
The rollup executes transactions on its own network and periodically publishes them along with the state root of the rollup network.
The state root is determined to be final by a protocol, often enforced by smart contracts on the anchoring blockchain, which may let the state roots be challenged or verify an accompanying validity proof.
While this core functionality is universal to existing rollups, these systems have introduced unique features as they vie for users and market dominance.
In this paper, we aim to classify ways in which these rollups differ in order to establish a common ground of understanding. 
We explore various dimensions in which these system can differ: familiarity, finality time, modularity, and maturity.
The result is a framework that can be used to understand and compare the properties of rollups.

\end{abstract}

\section{Introduction}\label{sec:introduction}
\subsection{Background}

A \emph{rollup} is a special kind of blockchain network.
In a rollup, transactions are executed and blocks are constructed, but users cannot be ultimately sure of the canonical rollup chain until it is \emph{finalised} (i.e., cannot be ignored as part of the chain) according to some conditions.
The rollup periodically publishes the transactions it processes, along with the state root (defined formally in Section~\ref{sec:preliminaries}) which summarizes the state of the rollup blockchain.
These state roots are considered final after some conditions are met, depending on the rollup.
The conditions are typically enforced by smart contracts\cite{szabo1997idea} that interpret and update the rollup state root on a different blockchain.

Rollups have become popular because their design enables a greater throughput than many blockchains that enforce their finality.
By executing transactions using their own consensus layer, a rollup network can execute more transactions than a blockchain that needs to replay each transaction on each node.
Though rollups eventually need consensus on which result is correct, this has been overcome by merely indicating the transactions to be processed (which is a relatively cheap operation) and establishing a protocol to consider the results valid unless proven otherwise, or by providing a mathematical proof that their execution was done correctly.
The Ethereum blockchain\cite{ethereum} can support only a handful of transactions per second\cite{eth15tps}, but rollups which post data to it and use it to enforce finality can process closer to 2000 transactions per second\cite{DBLP:journals/access/ThibaultSH22}.
This results in a better user experience and often reduces the fees users need to pay for transactions.
The blockchain on which the rollup-governing smart contracts are deployed is often called the \emph{layer one} blockchain and the rollup itself is the \emph{layer two}; the rollup can be seen to be built on top of the layer one.
This typically requires rollups to be built on top of blockchains that support smart contracts, like Ethereum.

The specifics of what data is published, the conditions necessary for finality, and many other properties, vary from rollup to rollup.
While some classes of rollups may aim to behave in similar fashion, others may not.
However, even those that aim to be similar may have subtle differences.
This difference may be intentional (e.g., by adding custom functionality) or unintentional (e.g., the rollup is too immature to be considered feature-complete).
Either way, this results in different risk models for different rollups.
Just as different blockchains have different properties\cite{GUO2022100067} and risks\cite{8371010}, different rollups come with different properties and risks.

However, even similar rollups do not necessarily use the same terminology.
Rollups are often commercial products, and there is value to differentiating one's product from a competitor's.
Each rollup introduces their own terms for unique features or renames common elements in order to stand out.
This problem is exacerbated by rollups which differ greatly and bare little resemblance to any others on the market.
The result is a mess of concepts that is not easy for researchers, let alone end-users, to comprehend.

This work aims to define and clarify modern rollup terms and concepts. 
In doing so, we build a framework for comparing rollups that is simple to understand and general enough to capture future rollup designs.

\subsection{Research Question and Method}
The proliferation of rollups as a means to overcome the shortcomings of general purpose blockchains have raised a number of concerns.
Rollups may appear similar but be very different, and they may be immature or outright insecure.
It is necessary to be able to easily compare rollups and their properties.

First, there may be a misconception that rollups are either very similar to each other, or to their underlying layer one. 
Neither is necessarily true, and while this may be apparent to the developers of these systems, it is not easy to discern for end-users or decentalized application (``dApp'')\cite{8466786} developers.
A change to the behaviour of a single virtual machine (VM) opcode or the inclusion of custom functionality may change the outcome of a transactions in ways that users are not prepared for.
These differences are often not well communicated to users.

Second, while modern layer one blockchains are well-established and often considered trustworthy, rollups are yet to be truly battle-tested.
Until this changes, there may be security and maturity considerations for these systems that are not present for layer one blockchains.
Some of these concerns are unavoidable with the rollup design (indeed, some may arise \emph{because} of it\cite{10.1145/3631310.3633493}), but others simply stem from rushing to market.
Despite several rollups operating mainnet networks, only a handful of specialized ones are feature-complete.
Rushed development of these systems put user funds and the trustworthiness of the ecosystem at risk.

These concerns may have different resolutions based on different properties of rollups.
In this work, we study the properties of real-world rollups and how their architecture characterizes them, and in doing so, we create an comparison framework for rollups.
We first establish a common language for these systems and their key components and actors.
Next, we survey existing rollups to establish how they are categorized and which properties are affected by each category.
From this survey we collect a list of modern rollup concepts, much of which is missing from existing academic literature.
Aiming to also provide a framework for the comparison of rollups and their properties, we propose four dimensions that can be used to contrast rollups based on their architecture and design.

\subsection{Results}
Our results are two-fold: first, we construct a collection of rollup concepts, and second, we provide a framework for comparing rollups and their properties.

The collection we construct includes various types of previously unstudied rollups, and we study modern features that are introduced by specific projects.
These have arisen from a variety of reasons, including technological interest and marketing requirements.
This collection of modern rollup concepts is enough to categorize existing rollups, though a single rollup may fit into several categories simultaneously.

Recognizing that a rollup may belong to several categories, we use these terms to build a framework for comparing rollups.
We build on four dimensions that intersect these categories and propose a framework that is powerful enough to illustrate the differences between rollup networks.
Since rollups will continue to evolve as they gain traction (or suffer exploitation), this framework cannot be considered complete.
However, the dimensions along which rollups are differentiated are unlikely to change, even in light of new research or technology.
Anything that requires another axis will deserve its own classification within blockchain networks.
Regardless, our provides a common language for these systems and a method to compare them.

This work goes hand-in-hand with other studies of blockchain systems, which have focused on layer one or standalone blockchains.
Taxonomies have been proposed\cite{DBLP:journals/ledger/TascaT19,DBLP:conf/icsa/XuWSZBBPR17, DBLP:journals/cluster/BallandiesDP22} for blockchain technologies, which are detailed but predate mainstream layer two solutions.
Some concerns are not relevant for most existing rollups (like consensus mechanisms) while others (like block storage size) might be just as relevant on these systems.
These works can be helpful in rollup settings for fine-grained analysis of particular features, but may also be too premature as these systems are undergoing active development.
These taxonomies leave out unique rollup concerns like data availability (see Section \ref{sec:modularity}) altogether.

Our work is similar to the work\cite{7890159} of Okada et al.~which aims to establish dimensions to classify layer one systems, but targets rollups instead.
This is complementary to surveys\cite{DBLP:journals/access/ThibaultSH22,DBLP:journals/jnca/GangwalGT23,DBLP:conf/fc/GudgeonMRMG20} of the more general notion of layer two scaling solutions for blockchains.
While helpful to define these systems and compare trade-offs, these surveys are too general and do not provide a concise set of properties to consider for rollup comparison.
Our work presents a more focused analysis aimed at helping others understand these prolific scaling solutions and their properties.

\section{Rollup Preliminaries}\label{sec:preliminaries}

In this section, we outline components of a rollup. 
Rollups are also known as \emph{commit-chains}\cite{cryptoeprint:2018/642} or \emph{validating bridges}\cite{cryptoeprint:2021/1589}.
Rollups require a system to order transactions, publish state roots, and a mechanism to verify and accept the state roots.
They also have a built-in bridge to link assets from the underlying layer one to the rollup itself and vice-versa.

In a rollup, transactions must be ordered to be executed.
Typically there is a single actor called the \emph{sequencer}, which performs this action.
It orders layer two transactions and commits to this ordering by publishing the transactions.
Often, this is posted as calldata to the layer one; however this is not always the case (see Section~\ref{sec:modularity}).
Data posted to Ethereum as calldata is intended to be used during a function call and then discarded; in particular, it is not accessible later on Ethereum unless the called function stores it.
As this data is part of a transaction, it is still included in the overall state of the blockchain, but because it is not explicitly stored, it is much cheaper than data which is stored on the blockchain.
The commitment to transactions to be executed is often called a \emph{batch} of transactions, as the data published usually contains multiple transactions.
This introduces a concept called \emph{soft finality}, whereby users who trust the rollup can consider transactions (and their results) in the batch as final.
However, others will wait for the conditions specified by the finality mechanism (see Section \ref{sec:finality-time}), which is sometimes called \emph{hard finality}\cite{10.1145/3631310.3633493}.
At the time of writing, most rollups have single sequencers operated by the rollup developers.
In the future, this role may be decentralized\cite{DBLP:journals/corr/abs-2310-03616} and either allow this role to rotate according to some rules, or introduce a consensus mechanism in order to order the layer two transactions.

Rollups also need to publish the \emph{state root} of the layer two blockchain.
A state root is the root of a state trie \cite{DBLP:journals/corr/abs-2108-05513} for the blockchain, typically a Merkle Patricia Trie (MPT)\cite{mpt-eth}.
This data structure contains a hash of all of the necessary data relevant to record the rollup state; for example, users' nonce values, smart contract code, and values for variables (including user balance).
The actor responsible for this action is often called the \emph{state proposer}.
This actor takes the ordered transactions and executes them (in the prescribed order) on the layer two, so that the state of the layer two is updated.
In practice, this is often the same actor as the sequencer, though this is not always true (see Section~\ref{sec:finality-time}).

A \emph{verifier} is necessary to determine which state roots are to be accepted and finalised.
This role may be explicit or implicit.
In some cases, explicit actors may be expected to replay state transitions (that is, the execution of the transactions from a given point) and verify that the same state root is obtained.
In other cases, a smart contract will verify a mathematical proof of the state transition and therefore there is no explicit actor.
The finality mechanism for the rollup determines the necessary verifier behaviour and type (see Section \ref{sec:finality-time}).

These terms for rollup architecture are not universally agreed upon.
Other works may have named these components differently\cite{idealescapehatches, cryptoeprint:2021/1589} but these components are critical to satisfy the requirements of a rollup which orders transactions, batches and publishes them, and ultimately publishes a state roots and finalises them.

Finally, a rollup has a \emph{built-in} or \emph{canonical bridge}.
A \emph{bridge} is a cross-chain communication protocol\cite{bridgehacksqspub}.
Bridges take assets or messages from one blockchain and create representations or copies of them on another blockchain.
The rollup's canonical bridge is necessary to bridge assets from the underlying layer one to the layer two itself.
In the case of a rollup, a layer one-to-layer two transaction is referred to as a \emph{deposit} transaction as it is a means of depositing layer one assets on the layer two.
The canonical representation of the native layer one asset (e.g., Ether) is bridged via deposit transactions, though the bridge may support other digital assets as well.
Note that the term ``deposit transaction'' has come to be overloaded, and may refer to any layer one-to-layer two transaction, even if an asset is not deposited as part of the transaction\cite{deposittx}.
A layer two-to-layer one transaction is often referred to as a \emph{withdrawal} transaction, as it is the main method by which digital assets are bridged off of the rollup.


Some have argued that a rollup network does not depend on smart contracts and a canonical bridge, as social consensus can determine which fork of the rollup is the valid one\cite{joncharbrollups}.
In such a case, the rollup simply requires its own network of nodes and does not necessarily require layer one smart contracts, and in particular, the bridge contracts; such a rollup is referred to as a \emph{sovereign rollup}\cite{sovereignrollups}.
While this may be true from a technical standpoint, a fork obtained by social consensus may not be collateralized: the consensus may need to override the layer one's smart contracts, which may not be able to release deposited funds in the canonical bridge in such a case.
This is important to mention as the goal of the paper is to discuss modern rollup terminology, and by rejecting this view, we establish that a rollup should have a canonical bridge.
Sovereign rollups are revisited in Section \ref{sec:modularity}.

We now introduce some conventions for this work.
For simplicity and concreteness, we will only consider Ethereum as a layer one in this work, but the results are not limited to rollups on Ethereum.
We will also refer to data and transactions that are executed by a (layer one) blockchain as \emph{on-chain}, while those which are stored or executed elsewhere as \emph{off-chain}.

\section{Dimensions}\label{sec:dimensions}

In this section, we introduce and justify the four dimensions for the framework outlined in Section \ref{sec:framework}.
First, we consider the \emph{finality time} of a rollup.
This property affects how composable a rollup is within an ecosystem and its user experience.
Second, we consider the \emph{familiarity} of a rollup, relative to both the underlying layer one and other rollups.
This property affects how transaction and dApp executions are expected to be completed by rollup users; unfamiliar rollups may introduce the risk of unexpected behaviour.
It also affects the user experience on the system, because users may need different assumptions or tools to interact with the rollup.
Third, we consider the \emph{modularity} of a rollup.
This property affects the trust assumptions of the rollup as there may be multiple systems that users must rely on, which in turn also affects the riskiness of the system.
Finally, we consider a rollup's \emph{maturity}.
The maturity of any complex network is a property that affects the system's robustness and reliability, as well as the likelihood of others to integrate the network into their own ecosystem.

\subsection{Finality Time}\label{sec:finality-time}
A rollup's finality time is a property that affects its level of composability and its user experience.
A rollup with a large finality time will have difficulty integrating into layer one dApps which need to be sure that the results of the state root will not be reverted, while those with smaller finality times will have an easier time being integrated.
In turn, the user experience for a rollup with a large finality delay may not be desirable.
There are two dominant methods for achieving finality in rollups: \emph{optimistically} and by using \emph{zero-knowledge proofs}.

Rollups that achieve finality optimistically are known as \emph{optimistic rollups}.
This name arises from the fact that a state proposer will submit a state root, and unless another actor can demonstrate that this state root is incorrect (given the ordered list of transactions and the state root to which they were applied) within a fixed time period, the state root will be accepted and considered final.
In other words, the state root is submitted with the optimistic belief that it will be accepted.
In a bug-free implementation operated by an honest state proposer, this state root should never be shown invalid.
The process for demonstrating demonstrating that the state root is known as a \emph{challenge} and can be initiated during the \emph{challenge period}, which is typically several days long\cite{7days-kelvin}.
The delay is required to be long in order to prevent attacks where a malicious state proposer can censor challenges on the layer one for the duration of the challenge period.

We now outline such a challenge.
Let $S$ be a state proposer being challenged by a verifier $C$.
As both $S$ and $C$ agree on the genesis state root for the rollup, they agree on a set of rollup blocks up to some point.
We can assume that $C$ agrees with $S$ on the state roots of all but the last one that $S$ proposed.
Then, since the rollup indicated which transactions were to be executed and their ordering, both $S$ and $C$ agree on which transactions should take them from the old state root to their proposed state root.
That is, $S$ has an execution trace $\rho$ which is the trace of state updates executed during the transactions in their view, and $C$ has an execution trace $\rho'$ which is a different trace that should result from executing the transactions in their view.
First, binary search is performed on the execution traces of $\rho$ and $\rho'$, aiming to find the largest common prefix that both $S$ and $C$ agree on.
This is possible as $S$ and $C$ must agree on a starting state root (the output from the previous block, or the genesis), or else $C$ should have initiated an earlier challenge by the assumption that these two actors agree on all prior blocks.
Then, the operation that is performed immediately after the agreed upon prefix within the transition consists only of a small operation; perhaps a variable update, or modification to the stack, depending on the level of ganularity.
This operation is small enough that it can be encoded and simulated on the underling layer one smart contracts for the rollup.
If the result does not match the next value in $\rho$, then $C$ has successfully challenged $S$.
Provided the value after the on-chain simulation matches the next one in $\rho'$, $C$ is often also allowed to propose this as the new state.
Note that in this challenge, $C$ is both an explicit verifier and a state proposer.
If the value matches the one in $\rho$, $C$'s challenge is unsuccessful.
In a successful challenge, the trace $\rho'$ is often called a \emph{fraud proof} as it proves that $S$ acted fraudulently.

To avoid denial-of-service attacks and misaligned incentives, state proposers in optimistic rollups are  bonded.
That is, they stake valuable cryptocurrency on the layer one, and the loser of any challenges forfeits their stake (typically to the other party).
Provided the stakes are large enough, this ensures that the original state proposer is incentivized to behave honestly and discourages would-be challengers from possibly slowing down the network by challenging every state root.

An alternative method for accepting state roots involves the use of \emph{zero-knowledge proofs}.
A zero-knowledge proof is generated by executing a computation inside a zero-knowledge proof system, like the one proposed by Groth in 2016\cite{Groth16}.
These systems use advanced algebraic and cryptographic techniques to transform the computation into a system of equations (or another representation), which allow the execution of the computation to proceed in such a way that another artifact, a \emph{validity proof}, is generated alongside any output.
The validity proof is a mathematical object which can only have been obtained if the specific computation, along with the given inputs, was computed correctly.
Moreover, the system dictates a method for verifying such a proof efficiently, often more efficiently than running the computation directly.
As a result of this efficiency, validity proofs can often be checked directly on smart contracts of modern blockchains, like Ethereum\cite{ethereum}.
This is true even in settings where the computation in question is to apply the state transition of several blockchain transactions\cite{cryptoeprint:2024/050}.
As a result, some rollups use this method to finalize state roots, and they are called \emph{zero-knowledge rollups}.
Zero-knowledge rollups can often be finalized in a matter of minutes: it may take a few minutes to construct the layer two block in the proof system, but once that is done, the corresponding validity proof can verified in seconds on the underlying layer one.
Accepting the state root then boils down to verifying the validity proof on-chain; provided that the verification succeeds, the state root is updated.

Regardless of the specific mechanics required to accept a new state root, the user experience and composability largely boils down to the delay between seeing a proposed layer two state root and the time it is considered final.
The composability of a zero-knowledge rollup is generally better: withdrawal transactions can be be processed as soon as their corresponding validity proof is verified.
In turn, this means that layer two-to-layer one transactions may immediately call layer one dApps in some designs.
Similarly, the (much quicker) certainty that a transaction will not be omitted from a state root (or have a different outcome because another transaction was fraudulent) provides a better user experience than one that technically (if not also pragmatically) requires a several day wait.
It is important to be able to distinguish and compare rollups according to the time required for their state roots to be considered final.

\subsection{Familiarity}\label{sec:familiarity}

As layer two scaling solutions, rollups rely on other block\-chains and aim to support users from them.
As a result, users who come from the layer one may expect some level of familiarity with the rollup network.
In the case of Ethereum, dApp developers want to be able to easily port their code over to the new chain without friction.
This typically includes the Solidity programming language\cite{solidity} and executing it on the Ethereum Virtual Machine (EVM).
Often this is the case and many Ethereum based rollups are EVM-compatible, meaning that they aim to support tools and languages that are supported by the EVM.
On these rollups, many Solidity smart contracts and other tools (e.g., wallets or software development kits) work out-of-the-box.
However, this is not always the case: for example, the Starknet zero-knowledge rollup\cite{starknet} uses a custom VM which does not resemble the EVM.
But even rollups which aim to be familiar for end users may still introduce unfamiliar differences.
These unfamiliar changes may be by design, due to technical limitations, or due to resource constraints.

Rollups may be unfamiliar to to users because of design decisions.
These decisions may be to support a different set of requirements or to stand out from competitors; we provide a few examples from rollups on the Ethereum blockchain.
Starknet chose to build a custom VM in order to easily leverage ZK-STARK technology, a quantum-resistant zero-knowledge proof system, which they had developed\cite{DBLP:journals/iacr/Ben-SassonBHR18}.
The Zircuit zero-knowledge rollup\cite{Zircuit}, which is EVM-compatible, will include so-called \emph{sequencer level security}.
This is a feature that generalizes the concept of Decentralized Finance (DeFi) circuit breakers\cite{sok-defi} that is built-in to the sequencer in order to guard against malicious transactions.
The Taiko zero-knowledge rollup\cite{taiko} has introduced other features still: a concept that they call \emph{boosted rollups}\cite{booster-rollups} for layer three rollups built on Taiko. 
Boosted rollups are ones that aim to shard transaction execution and storage, by allowing rollups  (on the same layer, operating in parallel) which implement new opcodes to delegate calls (i.e., execute them) in the context of the underlying layer's data directly.
These delegated calls allow the layer three rollups to make calls in the context of data stored on Taiko (i.e., layer two).

Other unfamiliar changes are more subtle and result from technical limitations.
Many rollups do not implement their layer one functionality exactly.
For example, the zkEVM zero-knowledge rollup by Polygon\cite{polygonzkevm} separates every transaction into its own block.
As a result, while Solidity smart contracts are supported on the rollup, the use of the \texttt{block.number} function within them may not be the same as on Ethereum.
This can be a problem if, for example, a dApp accrues interest based on the number of blocks that have elapsed: the interest earned will not be the same across different chains.
In fact, there are a number of differences between opcodes and built-in functions on rollups and their layer one; websites like RollupCodes\cite{rollupcodes} track these differences.
However, these differences may not be readily apparent to dApp developers and may be difficult to discern or comprehend for end-users.
As a result, these differences create unfamiliar user experiences that can be problematic.

Finally, the unfamiliar changes of a rollup may simply arise because of resource constraints.
In particular, underlying layer one networks are generally still evolving and adding features.
This means that rollups not only need to catch up to the state of the layer one as it existed when they started development, but also be ready to adjust to changes incorporated into the layer one during the development time.
For example, while the Optimism\cite{optimism} rollup aims to be EVM-compatible, it took several months for them to incorporate changes from the Shanghai hard fork of Ethereum\cite{shanghai-fork}.
This is not necessarily a technological challenge --- it is no harder than building a rollup in the first place --- but arose nonetheless because the underlying network artifacts evolved alongside the rollup itself.
In short, rollup developers often need to implement custom functionality while also tackling complex  dependencies introduced by the underlying layer one.
This is challenging given the size and complexity of these systems and requires many coordinated resources to do effectively.

Unfamiliarity affects a rollup's riskiness and user experience.
The assumptions that are safe to make by end-users and dApp developers may differ due to unfamiliar environments on these rollups.
Those assumptions are not always accurate due to subtle changes made by the rollup.
Similarly, the user experience is negatively affected when the transaction behaves differently on different chains or must be customized for a particular rollup.

\subsection{Modularity}\label{sec:modularity}
Rollups with a standalone sequencer that orders transactions, posts data to Ethereum, and derives state from Ethereum have been come to be known as \emph{monolithic} blockchains\cite{monolithicrollup}.
This is part because they largely consist of a single sequencer node, which is fills multiple roles (sequencer and state proposer) while also communicating to a single general-purpose blockchain as a layer one, which acts as the settlement, consensus, and data availability layer.
However, although the requirements of Section \ref{sec:preliminaries} are generally well accepted, the implementation details are not.
Some developers have chosen to move away from a monolithic model, either by breaking apart system components or using different blockchains (or other networks) to post data to or derive state from.
These novel approaches to each requirement have resulted in new terminology and concepts, and not all of these are universally accepted.
For example, the use of a zero-knowledge proof system for fast finality (Section \ref{sec:finality-time}) but a change to \emph{where} rollup data is posted (namely, off-chain) results in the term \emph{validium}\cite{validium, buterintypes}.
Not everyone agrees that these are truly rollups, as they have different trust assumptions. 
In this section, we explore how implementing each requirement in its own module affects the properties of a rollup (or rollup-like layer two).
We maintain that although some trust assumptions may change, any system that still satisfies the previously established requirements is worth considering as a rollup, especially when marketed as such.
These systems are kinds of \emph{modular} rollups, which may simply need more transparency.


Recall from Section \ref{sec:preliminaries} that sovereign rollups are those that are not smart contract based.
This also means that any settlement for the rollup is done on another blockchain.
As these rollups still maintain a layer two blockchain state, they must have distinct rollup nodes which order theses transactions on the layer two.
Standing in opposition to sovereign rollups, are so-called \emph{enshrined rollups}\cite{enshrinedrollups}.
These are rollups that are not smart contract based, but rather have their controlling logic embedded into their layer one code.
An imaginary enshrined zero-knowledge rollup on Ethereum may have the verification of its state roots be accomplished by Ethereum validators who are constructing the layer one blocks.
Sovereign rollups may also be \emph{accountable}\cite{DBLP:journals/corr/abs-2210-15017}, in that they can be implemented to determine which network nodes they are using for their settlement layer have tried to mislead them.
Trust assumptions for sovereign rollups differ from those of smart-contract based rollups: there may be different assumptions about the rollup network liveness, as it is not guaranteed by the underlying layer one.
These modular rollups change where settlement takes place.

Rollups may also be modular in other ways. 
They can change where layer two transactions are sequenced, or where data is posted.

A rollup can rely on another blockchain (namely, its layer one) for sequencing transactions; the result is a so-called \emph{based} rollup.
In this case, the data is posted is posted to the underlying layer one and the layer one itself is responsible for ordering the layer two transactions.
This can be done using its consensus mechanism: layer one block builders are also responsible for including the next rollup block as part of the layer one block.
As a result, these rollups are also sometimes called \emph{layer one sequenced}\cite{basedrollups}.
These have some desirable properties: if the underlying layer one is decentralized, the layer two is as well; there is no need for an escape hatch and generally have the same liveness guarantees as the layer one; and are cheap to operate, among other benefits.
This kind of rollup would require changes to the underling layer one infrastructure, however.
Moreover, the user experience is determined by how state transitions within a valid block are shown to be correct, and requires proposer-builder separation\cite{DBLP:conf/imc/HeimbachKTW23} on the underlying layer one.
Based rollups may be enshrined rollups, but they may differ if an enshrined rollup has a sequencer and uses the layer one, for example, to only verify state transitions as part of layer one blocks.

A rollup can also rely on another blockchain for publishing its transaction data.
This type of modularity is popularized by projects like Celestia\cite{celestia}, which provide different data availability networks that aim to be cheaper than Ethereum.
This results in different trust assumptions: if the data availability layer is online, these systems are similar to a standard rollup; if they are offline, funds may be lost.
Note that the lack of data does not let rollup operators (or anyone else) steal funds, but it may make it impossible to prove that transactions and funds existed.
Celestia notes that this solution is particularly suited for sovereign rollups, as it also allows them to easily change how blocks are included by removing the ties to a single underlying blockchain.

To conclude, rollup properties depend on the modularity of the system.
This modularity affects the implementation and tooling considerations (data may need to be read by from other sources or fewer services may need be built).
In turn, this affects risk by relying on many well-engineered components, for which the inoperability could result in problems.
In other cases, trust assumptions are not applicable to all systems and users need to be aware of other factors even when all of these systems operate normally.
This dimension provides a way to distinguish rollups that does not depend on either the familiarity of the execution environment or the finality time.

\subsection{Maturity}\label{sec:Maturity}
Finally, the properties of a rollup may differ based on its maturity.
Each component implementing a requirement of the system may fail to do so completely and this is especially true while the rollup is under development.
One of the core risks of a rollup is the ability to withdraw funds from the relevant smart contracts (assuming it is not a sovereign rollup) even in the event that the sequencer, state proposer, or other parties responsible for operating a rollup are not online.
Such functionality is often called an \emph{escape hatch}\cite{idealescapehatches}.
The website \href{www.l2beat.com}{L2Beat}\cite{l2beat} tracks various properties of rollups, and shows that only $18$ out of $38$ active rollups have this functionality for both sequencer and state proposer failure at the time of writing.
That means more than half of active rollup projects may lose user funds if their operators are offline or there is a bug in a critical component.

To understand this risk, Buterin proposed guidelines for measuring the maturity of a rollup\cite{rollupstages}.
These guidelines introduced a set of three milestones that rollups can progress through, which include things like feature completeness, but also diversification of key actors through multi-signature wallets\cite{9223287}.
Each milestone represents a stage in the rollup's development and maturity.
This result provides guidance for this dimension, but we will argue that it is insufficient on its own.
It is important to stress that rollups which agree along other dimensions but differ along this one should not be expected to behave the same way.

This difference in behavior can differ in both normal operation and during emergency situations.
During normal operation, a rollup in stage zero (which is the easiest milestone to hit) does not  require validity or fraud proofs.
As a result, a malicious operator could forge transactions in order to steal funds from the canonical bridge.
Rollups that have achieved stage one are not susceptible to the same risk, but may be subject to collusion between key actors who can upgrade the rollup or intervene in problematic situations.
Finally, rollups that achieve stage two have none of these problems but are also not required to have the aforementioned escape hatch functionality.
Indeed, the operators may only be able to upgrade contracts or take other emergency measures after a significant delay, but there is no guarantee of what those measures might be.
There is nothing preventing a malicious security council from enacting a malicious upgrade if given the opportunity.

The last dimension we consider in this work is therefore maturity.
This dimension may not appear entirely orthogonal to the others, as most rollups are changing as they mature.
For example, a rollup may aim to be familiar but still be in active development and have not yet had the time to implement difficult opcodes or pre-compiled contracts\cite{precompiled-contracts}.
However, this dimension is not the same even in this example: such a rollup may fail to also use multisig wallets to guard privileged roles.
That requirement is not something users are expected to be familiar with and therefore is disjoint from the familiarity dimension, but nonetheless affects a rollup's trustworthiness.
Similarly, these kinds of considerations are not related to component-wise modularity or finality time.
Immature rollups are also not as composable as their mature counterparts: their application programming interface (API) may not be stable, or they may not meet security requirements yet.
For these reasons, maturity deserves its own consideration when understanding rollup properties.

\section{Framework}\label{sec:framework}

We now turn our attention to combining the dimensions introduced in Section \ref{sec:dimensions} into a single framework.
Our aim is to introduce a framework that quickly conveys the important properties of a rollup based on these dimensions.

To do this, we chose to construct a visual tool that quickly conveys differences among these dimensions.
As the properties of a rollup depend on multiple dimensions, we need a presentation that can handle multiple axes, and we chose to use radar charts (a.k.a.~spider diagrams or Kiviat diagrams\cite{DBLP:journals/sigmetrics/Kolence73}).
This representation allows us to arrange the axes visually so that derived properties of the rollup are near to the dimensions of the rollup that are important to them.
The user experience of a rollup depends on the rollup's finality time and its familiarity.
The riskiness of a rollup depends on the rollup's familiarity to users and developers and its modularity.
The trust assumption necessary for end-users depends also on the modularity but cannot be entirely separated from its maturity.
Composability will not be considered high on immature rollups without battle-tested technology or those with long periods of finality.

The user experience of a rollup depends on the rollup's finality time and its familiarity.
A familiar rollup will allow existing tools to work out of the box.
Users may be able to use the same wallets and developers can use the same smart contract language and infrastructure to deploy dApps.
A longer finality time will negatively affect user experience for cross-chain interactions.

Technical risks for end-users and dApp developers depend on the rollups familiarity and modularity.
Unfamiliar rollups may change how opcodes are implemented or use an entirely different VM; these may mean that otherwise trustworthy code behaves in unexpected ways.
Similarly, modular rollups which depend on other services (like different data availability solutions) add dependency risk.
These services may not always be online when needed by the rollup, and introduce other sources of errors.

A related concept, the rollup's trustworthiness, depends on assumptions introduced by modularity and the maturity of the project.
As argued in Section \ref{sec:modularity}, modular systems may introduce different trust assumptions.
This is different than technical risk in that these assumptions hold even when all systems are implemented correctly.
The maturity of a rollup also impacts its trustworthiness: recall that stage zero rollups may not be trustworthy at all if they fail to have a working proof or challenge system.

Composability depends on the finality time and the maturity of the project.
Many applications, especially those in DeFi, cannot accept withdrawn funds directly from rollups before their finality period has elapsed; those funds might not be valid.
Rollups with shorter finality times can be more closely coupled into dApp ecosystems as they do not slow these ecosystems down.
dApp developers will not be eager to accept funds on systems that are not battle-tested, as it may negatively impact their own dApp's operation.

By pairing up these axes, the area between these two axes can be used as visual signal to readers.
By convention, we can choose either a larger area or a smaller area to be consider more desirable.
We chose smaller areas, so that rollups with a large visual presence stands out.
As a result, some axes are renamed to their negation: we use an axis for ``unfamiliarity'' rather than ``familiarity'' and ``immaturity'' over ``maturity''.
The result is shown in Figure \ref{fig:framework-example} (a).

\begin{figure}[bt]
    \centering
     \subfloat[][Understanding the relation of the dimensions used in the radar chart. The area between the origin and the position of two neighbouring axes impacts a particular property of a rollup. For example, the unfamiliarity of a rollup, along with the finality time of the rollup, affect the user experience of the rollup. In general, smaller areas are better.]{
    \begin{tikzpicture}
\tkzKiviatDiagram[scale=0.35,
        label distance=1cm,
        radial  = 4,
        gap     = 1,  
        lattice = 5]{, \small{Unfamiliarity}, , \small{Immaturity}}
\draw (-7.75,0) node {\small{Modularity}};
\draw (8,0) node {\small{Finality Time}};     
\tkzKiviatLine[thick,color=red,mark=none,
               fill=red!20,opacity=.5](0,4,4,0)
\draw (5.25,4) node {\color{blue}\small{Negative}};               
\draw (5.25,3) node {\color{blue}\small{User Experience}};
\draw (5.25,-3) node {\color{olive}\small{Not Composable}};   
\draw (-5.25,-3) node {\color{orange}\small{Untrustworthiness}};
\draw (-5.25,3) node {\color{red}\small{Riskiness}};
\tkzKiviatLine[thick,color=blue,mark=none,
               fill=blue!20,opacity=.5](4,4,0,0)
\tkzKiviatLine[thick,color=orange,mark=none,
               fill=orange!20,opacity=.5](0,0,4,4)
\tkzKiviatLine[thick,color=olive,mark=none,
               fill=olive!20,opacity=.5](4,0,0,4)               
\end{tikzpicture}
    }%
    \subfloat[][Comparing two rollups using the radar chart explained in (a). Starknet is presented in teal while Optimism is in brown. Optimism is considered monolithic, so it does not have a presence on the modularity axis, but Starknet is a volition, where users can choose where their data is posted. However, as a zero-knowledge rollup, Starknet is closer to the origin along the finality time axis. Starknet has been in development for several years (so it is comparable to Optimism in maturity), but uses an entirely different VM, which makes it unfamiliar to Ethereum users.]{
    \begin{tikzpicture}
\tkzKiviatDiagram[scale=0.35,
        label distance=1cm,
        radial  = 4,
        gap     = 1,  
        lattice = 5]{, \small{Unfamiliarity}, , \small{Immaturity}}
\draw (-7.75,0) node {\small{Modularity}};
\draw (8,0) node {\small{Finality Time}};     
\tkzKiviatLine[thick,color=teal,mark=ball,fill=teal!20,opacity=.5](1,5,1,2) 
\tkzKiviatLine[thick,color=brown,mark=ball,fill=brown!20,opacity=.5](4,1,0,2) 
\draw (7.75,6.5) node {\small{\bf{Legend}}};
\draw (7.75,5) node[rectangle,draw,color=teal,fill=teal!20] {\color{teal}\small{Starknet}};
\draw (7.75,3) node[rectangle,draw,color=brown,fill=brown!20,] {\color{brown}\small{Optimism}};
\end{tikzpicture}
    }
    \caption{The visual framework in presented in Section \ref{sec:framework}. A guide to understanding its construction is presented in (a), while an example of its use is shown in (b).}
    \label{fig:framework-example}
\end{figure}

We can illustrate the effectiveness of this model by evaluating existing rollups.
Figure \ref{fig:framework-example} (b) shows a representation of two rollups, Starknet and Optimism.
From the figure, it is clear what the different properties of the rollups are: both have the same amount of maturity, but Starknet's finality time is much smaller as a zero-knowledge rollup.
Moreover, the fact that Starknet is not EVM-compatible is immediately clear.
Starknet is not on the origin for the modularity axis, since users can actually choose where their data is posted for the data availability (this makes it a so-called \emph{volition}\cite{starknet-volition}).
Applying this framework to all of the rollups on L2Beat.com\cite{l2beat} is left as future work.
\section{Conclusion}\label{sec:conclusion}

In this work, we have explored the categorization and labelling of rollups and presented a way to compare these complex systems.
After establishing a common ground for terminology, we explored categories of rollups which heretofore have not been considered in academic literature (like based, boosted, and modular rollups).
This exploration enabled a concise list of dimensions along which rollups can be compared, regardless of implementation techniques or custom features.
In turn these properties of familiarity, finality, modularity, and maturity can be used to discuss various pragmatic aspects of rollups.
These are the user experience of the rollup, the sources of risk, the assumptions required by users, and the composability of the system.
These dimensions can also be visualized, which in turn visualizes these properties in an easy to read graphical setting.
Just as rollups are not completely developed, the ideas and framework presented in this work is incomplete.
However, it can be used as a starting point on which to build more fine-grained analysis, explore changes to rollups not captured by terms in this document, and build an understanding of these complex systems.


\ledgernotes





\newpage


\bibliographystyle{unsrt}
\bibliography{ref}




\thispagestyle{pagelast}

\end{document}